\documentclass[
prl,
aps,
reprint,
superscriptaddress,
amsmath,amssymb,
]{revtex4-2}
\newcommand{\beq}{\begin{eqnarray}} % Shortcut for equation arrays
\newcommand{\eeq}{\end{eqnarray}}

\usepackage{silence}
\usepackage{graphicx} % For including images
\graphicspath{{./Images/}} % Specifies the directory where pictures are stored
\usepackage{dcolumn}% Align table columns on decimal point
\usepackage{bm}% bold math
\usepackage{nameref} % <-- 手动加载
\usepackage{hyperref}
\hypersetup{
    colorlinks=true,  % 启用颜色链接
    linkcolor=blue,   % 内部链接（如~\ref）的颜色
    citecolor=blue,   % 引用链接的颜色
    urlcolor=blue,     % URL 链接的颜色
    pdfencoding=auto, % 自动处理编码
    psdextra,         % 支持额外的 PDF 字符串
}

\begin{document}

%\preprint{}

%\title{Observation of Tensor-Driven High-Momentum Neutrons in $^{16}$O through Zero-Degree ($p$,$d$) Reactions}
\title{Observation of Tensor-Driven High-Momentum Neutrons in $^{16}$O via ($p$,$d$) Reactions and Zero-Degree Deuteron Momentum Spectroscopy}

% \author{}
%\thanks{}
% \affiliation{}


\author{X.~Wang}
\affiliation{School of Physics, Beihang University, Beijing 100191, China}
\affiliation{Research Center for Nuclear Physics, Osaka University, Ibaraki, Osaka 567-0047, Japan}
\author{H.~J.~Ong}
\email{onghjin@impcas.ac.cn}
\affiliation{State Key Laboratory of Heavy Ion Science and Technology, Institute of Modern Physics, Chinese Academy of Sciences, Lanzhou 730000, China}
\affiliation{School of Nuclear Science and Technology, University of Chinese Academy of Sciences, Beijing 100049, China}
\affiliation{Joint Department for Nuclear Physics, Lanzhou University and Institute of Modern Physics, Chinese Academy of Sciences, Lanzhou 730000, China}
\affiliation{Research Center for Nuclear Physics, Osaka University, Ibaraki, Osaka 567-0047, Japan}
\affiliation{Nishina Center for Accelerator-Based Science, RIKEN, 2-1 Hirosawa, Wako, 351-0198 Saitama, Japan}
\author{S.~Terashima}
\email{tera@impcas.ac.cn}
\affiliation{State Key Laboratory of Heavy Ion Science and Technology, Institute of Modern Physics, Chinese Academy of Sciences, Lanzhou 730000, China}
\affiliation{School of Physics, Beihang University, Beijing 100191, China}
\author{I.~Tanihata}
\affiliation{Research Center for Nuclear Physics, Osaka University, Ibaraki, Osaka 567-0047, Japan}
\affiliation{School of Physics, Beihang University, Beijing 100191, China}
\author{Y.~K.~Tanaka}
\affiliation{High Energy Nuclear Physics Laboratory, RIKEN, 2-1 Hirosawa, Wako, 351–0198 Saitama, Japan}
\author{N.~Aoi}
\altaffiliation{Present address: Center for Nuclear Study, University of Tokyo, Tokyo 113-0033, Japan}
\affiliation{Research Center for Nuclear Physics, Osaka University, Ibaraki, Osaka 567-0047, Japan}
\author{Y.~Ayyad}
\affiliation{Universidade de Santiago de Compostela, 15782 Santiago de Compostela, Spain}
\author{J.~Benlliure}
\affiliation{Universidade de Santiago de Compostela, 15782 Santiago de Compostela, Spain}
\author{F.~Farinon}
\affiliation{MedAustron, Marie Curie-Strasse 5, 2700 Wiener Neustadt, Austria}
\author{H.~Fujioka}
\affiliation{School of Science, Institute of Science Tokyo, 2-12-1 Ookayama, Meguro, Tokyo 152-8551, Japan}
\author{H.~Geissel}
\altaffiliation{Deceased.}
\affiliation{Universität Giessen, Heinrich-Buff-Ring 16, 35392 Giessen, Germany}
\affiliation{GSI Helmholtzzentrum für Schwerionenforschung GmbH, Planckstraße 1, 64291 Darmstadt, Germany}
\author{J.~Gellanki}
\affiliation{KVI-CART, University of Groningen, Zernikelaan 25, 9747 AA Groningen, The Netherlands}
\author{C.~L.~Guo}
\affiliation{Department of Radiation Oncology, National Cancer Center/National Clinical Research Center for Cancer/Cancer Hospital, Chinese Academy of Medical Sciences and Peking Union Medical College, Beijing, 100021, China}
\author{E.~Haettner}
\affiliation{GSI Helmholtzzentrum für Schwerionenforschung GmbH, Planckstraße 1, 64291 Darmstadt, Germany}
\author{W.~L.~Hai}
\affiliation{School of Physics, Beihang University, Beijing 100191, China}
\author{M.~N.~Harakeh}
\affiliation{GSI Helmholtzzentrum für Schwerionenforschung GmbH, Planckstraße 1, 64291 Darmstadt, Germany}
\affiliation{ESRIG, University of Groningen, Zernikelaan 25, 9747 AA Groningen, The Netherlands}
\author{C.~Hornung}
\affiliation{Universität Giessen, Heinrich-Buff-Ring 16, 35392 Giessen, Germany}
\author{K.~Itahashi}
\affiliation{Nishina Center for Accelerator-Based Science, RIKEN, 2-1 Hirosawa, Wako, 351-0198 Saitama, Japan}
\author{R.~Janik}
\altaffiliation{Deceased.}
\affiliation{Comenius University Bratislava, Mlynská dolina, 842 48 Bratislava, Slovakia}
\author{N.~Kalantar-Nayestanaki}
\affiliation{ESRIG, University of Groningen, Zernikelaan 25, 9747 AA Groningen, The Netherlands}
\author{R.~Knöbel}
\affiliation{Universität Giessen, Heinrich-Buff-Ring 16, 35392 Giessen, Germany}
\affiliation{GSI Helmholtzzentrum für Schwerionenforschung GmbH, Planckstraße 1, 64291 Darmstadt, Germany}
\author{N.~Kurz}
\affiliation{GSI Helmholtzzentrum für Schwerionenforschung GmbH, Planckstraße 1, 64291 Darmstadt, Germany}
\author{K.~Miki}
\affiliation{Tohoku University Department of Physics, Graduate School of Science and Faculty of Science Aza-Aoba 6-3, Aramaki, Aoba-ku, Sendai, 980-8578, Japan}
\author{I.~Mukha}
\affiliation{GSI Helmholtzzentrum für Schwerionenforschung GmbH, Planckstraße 1, 64291 Darmstadt, Germany}
\author{T.~Myo}
\affiliation{General Education, Faculty of Engineering, Osaka Institute of Technology, Osaka 535-8585, Japan}
\affiliation{Research Center for Nuclear Physics, Osaka University, Ibaraki, Osaka 567-0047, Japan}
\author{T.~Nishi}
\affiliation{Nishina Center for Accelerator-Based Science, RIKEN, 2-1 Hirosawa, Wako, 351-0198 Saitama, Japan}
\author{D.~Y.~Pang}
\affiliation{School of Physics, Beihang University, Beijing 100191, China}
\author{S.~Pietri}
\affiliation{GSI Helmholtzzentrum für Schwerionenforschung GmbH, Planckstraße 1, 64291 Darmstadt, Germany}
\author{A.~Prochazka}
\affiliation{MedAustron, Marie Curie-Strasse 5, 2700 Wiener Neustadt, Austria}
\author{C.~Rappold}
\affiliation{Instituto de Estructura de la Materia, CSIC, Madrid, Spain}
\affiliation{GSI Helmholtzzentrum für Schwerionenforschung GmbH, Planckstraße 1, 64291 Darmstadt, Germany}
\author{M.~P.~Reiter}
\affiliation{Universität Giessen, Heinrich-Buff-Ring 16, 35392 Giessen, Germany}
\author{J.~L.~Rodríguez-Sánchez}
\altaffiliation{Present address: CITENI, Campus Industrial de Ferrol, Universidade da Coruna, 15403, Ferrol, Spain}
\affiliation{Universidade de Santiago de Compostela, 15782 Santiago de Compostela, Spain}
\author{C.~Scheidenberger}
\affiliation{GSI Helmholtzzentrum für Schwerionenforschung GmbH, Planckstraße 1, 64291 Darmstadt, Germany}
\affiliation{Universität Giessen, Heinrich-Buff-Ring 16, 35392 Giessen, Germany}
\affiliation{Helmholtz Forschungsakademie Hessen fuer FAIR (HFHF), GSI Helmholtzzentrum für Schwerionenforschung, Campus Giessen, 35392 Giessen, Germany}
\author{H.~Simon}
\affiliation{GSI Helmholtzzentrum für Schwerionenforschung GmbH, Planckstraße 1, 64291 Darmstadt, Germany}
\author{B.~Sitar}
\affiliation{Comenius University Bratislava, Mlynská dolina, 842 48 Bratislava, Slovakia}
\author{P.~Strmen}
\altaffiliation{Deceased.}
\affiliation{Comenius University Bratislava, Mlynská dolina, 842 48 Bratislava, Slovakia}
\author{B.~H.~Sun}
\affiliation{School of Physics, Beihang University, Beijing 100191, China}
\author{K.~Suzuki}
\affiliation{Research Center for Nuclear Physics, Osaka University, Ibaraki, Osaka 567-0047, Japan}
\affiliation{Stefan-Meyer-Institut für Subatomare Physik, Boltzmangasse 3, 1090 Vienna, Austria}
\author{I.~Szarka}
\affiliation{Comenius University Bratislava, Mlynská dolina, 842 48 Bratislava, Slovakia}
\author{H.~Toki}
\affiliation{Research Center for Nuclear Physics, Osaka University, Ibaraki, Osaka 567-0047, Japan}
\author{H.~Weick}
\affiliation{GSI Helmholtzzentrum für Schwerionenforschung GmbH, Planckstraße 1, 64291 Darmstadt, Germany}
\author{E.~Widmann}
\affiliation{Stefan-Meyer-Institut für Subatomare Physik, Boltzmangasse 3, 1090 Vienna, Austria}
\author{J.~S.~Winfield}
\altaffiliation{Deceased.}
\affiliation{GSI Helmholtzzentrum für Schwerionenforschung GmbH, Planckstraße 1, 64291 Darmstadt, Germany}
\author{X.~D~Xu}
\affiliation{State Key Laboratory of Heavy Ion Science and Technology, Institute of Modern Physics, Chinese Academy of Sciences, Lanzhou 730000, China}
\affiliation{School of Nuclear Science and Technology, University of Chinese Academy of Sciences, Beijing 100049, China}
\affiliation{GSI Helmholtzzentrum für Schwerionenforschung GmbH, Planckstraße 1, 64291 Darmstadt, Germany}
\author{J.~W.~Zhao}
\affiliation{School of Physics, Beihang University, Beijing 100191, China}
\collaboration{Tensor-Force/Super-FRS Experiment Collaboration}
\noaffiliation

\date{\today}

\begin{abstract}
The $^{16}$O($p$,$d$)$^{15}$O reaction has been studied at 0$^\circ$ using 403-, 604-, 907- and 1209-MeV protons, comparing cross sections populating positive- and negative-parity states in $^{15}$O. Transitions to positive-parity states exhibit strong enhancement due to high-momentum neutrons, while negative-parity transitions show much smaller effects. The cross-section ratio between positive- and negative-parity states rises sharply with momentum transfer, matching theoretical predictions that include tensor interactions, particularly the peak near 2 fm$^{-1}$. These results highlight 0$^\circ$ neutron-pickup reactions as a sensitive probe for tensor-driven high-momentum components, paving the way for studies in exotic nuclei via radioactive beams.
\end{abstract}

%\keywords{}

\maketitle

% \begingroup
% \renewcommand\thefootnote{\ddag}
% \footnotetext{Deceased.}
% \endgroup
% \begingroup
% \renewcommand\thefootnote{}
% \footnotetext{\textsuperscript{\ddag}Deceased.}
% \endgroup

% Nucleon-nucleon correlations
Tensor and short-range repulsive interactions are pivotal in nuclear physics and nuclear astrophysics, driving phenomena beyond mean-field descriptions~\cite{Hen2017}. 
In atomic nuclei, the tensor interactions -- arising mainly from pion exchange -- generate strongly correlated nucleon-nucleon pairs, modify spin-orbit splitting, and reshape magic numbers in exotic nuclei~\cite{Otsuka2005,Brown2006}. 
Together with the short-range correlations (SRCs) -- induced by the strong repulsive interactions between nucleons -- the tensor correlations generate configuration mixing via two-particle-two-hole (2p-2h) excitations, producing high-momentum nucleon pairs~\cite{Subedi2008,Wiringa2014}. 
These correlations critically impact spectral functions, quench single-particle strengths~\cite{Lapikas1993}, and underpin nuclear saturation and the equation of state (EoS)~\cite{Dickhoff2004}. 
Their astrophysical implications are profound: affecting the stiffness of the EoS, determining stellar radii and tidal deformabilities in neutron stars, and impacting neutrino opacities in supernovae and the synthesis of heavy elements via the r-process nucleosynthesis~\cite{Gandolfi2015,Abbott2018,Roberts2012,Horowitz2019}. 

% Tensor correlations and 2p-2h
The discovery of $S$-$D$ wave mixing in the deuteron~\cite{Schwinger1939,Rarita1941}, which explained its quadrupole moment~\cite{Kellog1939}, provided clear evidence of mixing of high-momentum nucleons in nuclei. 
Since then, mixing of nucleons with high momentum or in high-excitation orbitals and its effect on nuclear structure and response have been studied. This includes quenching of Gamow-Teller transition strengths and isovector magnetic moments~\cite{Bertsch1982,Towner1979,Wakasa1997}. 
Although tensor interactions are expected to play a key role, substantial contributions from exchange currents and $\Delta$-hole effects hinder isolating and extracting their distinct effect.
The isoscalar magnetic moments exhibit distinctive behavior, and are %unaffected by the effects observed in the isovector part due to angular-momentum conservation
well explained by incorporating 2p-2h excitations induced by tensor interactions extending to very high-energy orbitals~\cite{Shimizu1974,Towner1983,Arima1985}. 
%A model incorporating 2p-2h excitations induced by tensor interactions extending to very high-energy orbitals -- corresponding to $D$-wave mixing in deuterons -- successfully explained the isoscalar magnetic moments~\cite{Shimizu1974,Towner1983}. 
A similar anomaly observed in the isoscalar $M1$ transition has also been attributed to tensor interactions~\cite{Matsubara2015}. 
However, because these studies do not explicitly link high-excitation orbitals to high-momentum components,
%For example, in Refs.~\cite{Bertsch1982,Wakasa1997}, excitation energies up to 50 MeV were considered, but this corresponds only to momenta up to 0.45 fm$^{-1}$. 
the impact of high-momentum components on these observables is unclear.
While \textit{ab initio} calculations reveal high-momentum nucleons in nuclei~\cite{Wiringa2014} and the crucial role of tensor interactions in reproducing spectra of light nuclei with $A \leq$12~\cite{Pieper2001}, their effect on individual states remains unresolved. Models like the tensor-optimized shell model (TOSM)~\cite{Myo2009} and antisymmetrized molecular dynamics~\cite{Myo2017}, which include full 2p-2h configurations show promise but face challenges when applied to $p$-shell nuclei.

Experimentally, quasi-free electron~\cite{Benhar1993,Blomqvist1998,Subedi2008,Hen2017,Schmidt2020} and proton~\cite{Mardor1998,Tang2003,Patsyuk2021} scatterings have been extensively used to study high-momentum nucleons in nuclei, revealing short-range correlated nucleon pairs with momenta above the Fermi momentum. 
These reactions measure nucleon momentum distributions, showing that a significant fraction of nucleons form strongly correlated pairs. Both electron- and proton-induced knockout studies consistently highlight the importance of tensor correlations and SRCs in understanding nuclear structure and nucleon dynamics at short distances.
While electron scattering experiments on exotic nuclei~\cite{Suda2017} remain extremely challenging, proton-induced knockout reactions are often limited by low statistics as they require the simultaneous detection of multiple particles, thus hindering understanding of tensor and SRC effects on individual states.

% one-neutron-transfer reactions
A recent $^{16}$O($p$,$pd$) experiment employing neutron-pickup kinematics revealed strong cross-section enhancement for the lowest $1^+$ ($S=$1, $T=$0) state in $^{14}$N at large momentum transfers, directly indicating tensor-driven high-momentum neutron-proton pairs~\cite{Terashima2018}. 
$^{16}$O($p$,$d$) reactions were also studied at proton energies of 198, 295, and 392 MeV and a laboratory angle of 10$^\circ$~\cite{Ong2013}.
Two cross-section ratios were examined: $R_+$ for the positive-parity state ($1/2^+$ or $5/2^+$) and $R_-$ for the negative-parity state ($3/2^-$), both relative to the ground state. Their dependence on momentum transfer was compared with previous data~\cite{Snelgrove1969,Roos1975,Lee1967,Abegg1989,Smith1984}. 
While $R_-$ increases slowly and saturates above 1 fm$^{-1}$, $R_+$ rises rapidly, indicating possible role of nucleon-nucleon correlations, especially those from tensor interactions. 
However, linking these cross sections to neutron momentum distribution in $^{16}$O remains nontrivial.

% BA and unique importance of 0-deg
In ($p$,$d$) reactions with deuterons emitted near 0$^\circ$, the reaction is primarily driven by the pickup of a single neutron from the target nucleus~\cite{Kobushkin1986}.
Under the Born Approximation (BA), the cross section is proportional to the momentum distribution (MD) of the picked up neutrons in the target nucleus~\cite{Chew1950}. 
Although in reality the cross sections are modified by scatterings and distortion of the incident and emitted particles by the nuclear potentials, the modification has been shown to be minimum at 0$^\circ$~\cite{Berthet1982,Noble1974}. 
Therefore, the cross-section ratio at 0$^\circ$ reflects the MD differences of neutrons picked up from different orbitals.
For a reaction involving shell-model like states such as the negative-parity states in $^{15}$O, the Distorted-Wave Born Approximation (DWBA) calculation can reproduce the differential cross sections reasonably well. 
While DWBA successfully describes the positive-parity-state data at low-momentum transfer~\cite{Snelgrove1969}, neither DWBA nor coupled-channels Born approximation can reproduce the high-momentum-transfer results~\cite{Ong2013, Smith1984} -- expected to be due to missing high-momentum components in the wave functions.
The $^{16}$O($p$,$d$) reaction data at 0$^\circ$ therefore offer unprecedented opportunity to resolve the discrepancies.

Here, we present the first measurement of the $^{16}$O($p$,$d$) differential cross sections at 0$^\circ$ using four incident beam energies ranging from 400 to 1200 MeV. Transitions to the negative-parity states of $^{15}$O (the 1/2$^-$ ground state and the 3/2$^-$ state at 6.18 MeV), as well as the positive-parity 5/2$^+$ state at 5.24 MeV, are reported. 

%\section{Experiment} 
The experiment was performed at GSI Helmholtzzentrum für Schwerionenforschung in Darmstadt, alongside the experiment reported in Refs.~\cite{Tanaka2016,Tanaka18}.
Proton beams from the SIS-18 synchrotron with energies of 403, 604, 907, and 1209 MeV were used to cover momentum transfer around 2 fm$^{-1}$. 
%The typical beam intensity was 10$^{10}$ particles/spill, and the typical beam spot size at the target was 2 and 3 mm in horizontal and vertical directions, respectively. 
The proton beam bombarded a strip polyoxymethylene (POM; (CH$_2$O)$_n$) target with 1-mm horizontal width placed at the target station (TA) of the Fragment Separator (FRS)~\cite{Geissel1992}. 
We used 100(3)- and 308(6)-mg/cm$^2$-thick POM targets for the measurements with the two lower- and two higher-energy proton beams, respectively. 
Other targets included 107(4)- and 327(1)-mg/cm$^2$ carbon for $^{12}$C background subtraction, and 1027(2)-mg/cm$^2$ deuterated polyethylene (CD$_2$) for transmission efficiency calibration.
%All targets except CD$_2$ had 1-mm horizontal width to constrain beam dimensions.
Deuterons were momentum analyzed by the FRS in four added dispersive stages mode~\cite{FRSspec}, and detected at the final focal plane (S4). 
%In addition to the multi-wire drift chambers (MWDCs) and the SC41 plastic scintillator described in Refs.~\cite{Tanaka2016,Tanaka18}, we installed three additional plastic scintillators -- each with an identical active area of 200 $\times$ 50 mm$^2$ -- downstream of SC41 to measure the time-of-flight (TOF) of light charged particles.
The momentum dispersion at S4 was about 19 cm/\%. 
Distinct magnetic rigidity settings were applied for the C and POM targets to cover states of interest at 403 MeV; a single rigidity setting sufficed for both targets at other energies.
Detailed descriptions of the experiment conditions and detector setup are provided in Refs.~\cite{Tanaka2016,Tanaka18} and elsewhere~\cite{Wang}.

% Data analysis
\begin{figure}
\includegraphics[width=0.49\textwidth]{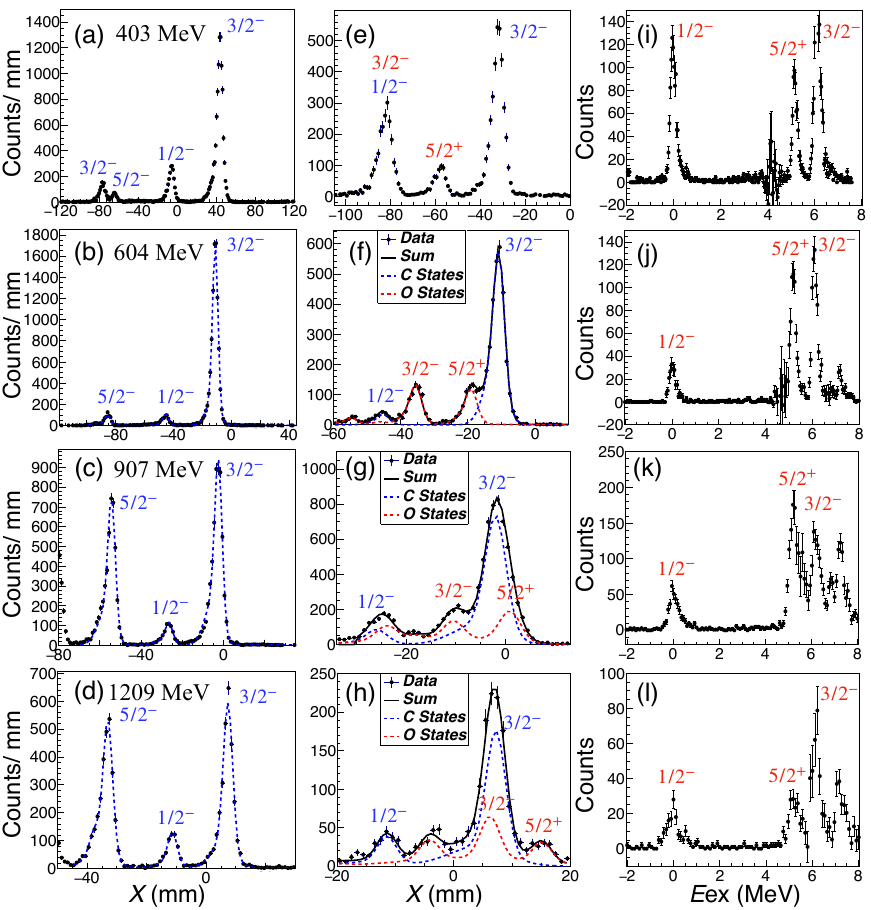}%
\caption{Position spectra of carbon (a)--(d) and POM (e)--(h), and excitation-energy spectra of $^{15}$O (i)--(l). 
Top panels show the data at 403 MeV (top), where discrete peaks are integrated individually without fitting.
Lower panels show the data at higher incident energies: 
blue-dashed and red-dashed lines correspond to the $^{11}$C- and $^{15}$O-state fits, respectively; 
black-solid lines represent the overall fits.
}
\label{fig:spec}
\end{figure}
%Since deuterons emitted at 0$^\circ$ from the ($p$,$d$) reactions have more than 25$\%$ higher magnetic rigidity than the proton beams, no protons could reach S4. 
%Primary triton backgrounds produced at target or beamline were removed event-by-event by the particle identification using the TOF and energy loss in plastic scintillators.
The horizontal position ($X$) distribution of the observed deuterons for the C and POM target measurement  are shown in Fig.~\ref{fig:spec}~(a)--(d) and (e)--(h), respectively.
To achieve the best resolution, we applied corrections accounting for optical system aberrations and the shift of beam energy correlated with extraction time. At 403 MeV, except for the 6.18-MeV state in $^{15}$O, which fully overlaps with the 2.00-MeV (1/2$^-$) state in $^{11}$C, other $^{15}$O states of interest are well resolved. 
For the data between 604 and 1209 MeV, the C-position spectra were first fitted with double-Gaussian functions before being subtracted from the POM spectra.
The relative height, peak-position difference and two widths defining the shape of the Gaussian functions were taken as free parameters but kept constant across all states.
Focusing on the overlapping region, we fitted the ground state (g.s.; $3/2^-$) and the 2.00($1/2^-$)-, 4.32($5/2^-$)-, and 4.80($3/2^-$)-MeV excited states of $^{11}$C, as shown by the blue-dashed lines in Fig.~\ref{fig:spec}(b)--(d).
Keeping the $^{11}$C population ratios, the $^{15}$O states were then fitted together with the $^{11}$C states in the POM spectra, as shown by the red-dashed and blue-dashed lines, respectively, in Fig.~\ref{fig:spec}(f)--(h).
The black-solid lines represent the minimum-$\chi^2$ fit to the POM spectra, determining the common shape parameters for both the $^{15}$O and $^{11}$C states, as well as the strength of the $^{11}$C states.
Several known low-lying excited states in $^{15}$O: 5.24(5/2$^+$), 6.18(3/2$^-$), 6.79(3/2$^+$), 6.86(5/2$^+$), 7.28(7/2$^+$), 8.92(5/2$^+$), 9.66(3/2$^-$) and 10.30(5/2$^+$) MeV were included in the fitting, while the ground state was well separated and thus treated separately. 
We did not consider the 1/2$^+$ excited state at 5.18 MeV since the previous work~\cite{Snelgrove1969} and conventional shell-model calculations indicate its contribution is negligible. 
%, and a subsequent high-resolution 0$^\circ$ $^{16}$O(p,d) experiment at 392-MeV proton energy [reference]. 
Subtracting the fitted C spectra from the POM spectra, we obtained the $^{15}$O position spectra. 
The deuteron momentum spectra were obtained from their linearly correlated $X$ positions, allowing subsequent determination of $^{15}$O excitation-energy spectra. 
The resultant $^{15}$O excitation-energy spectra, with indicators for states of interest, are presented in Fig.~\ref{fig:spec}(i)--(l), where corrections for the FRS transmission efficiency have been applied.

\begin{figure}
\includegraphics[width=0.4\textwidth]{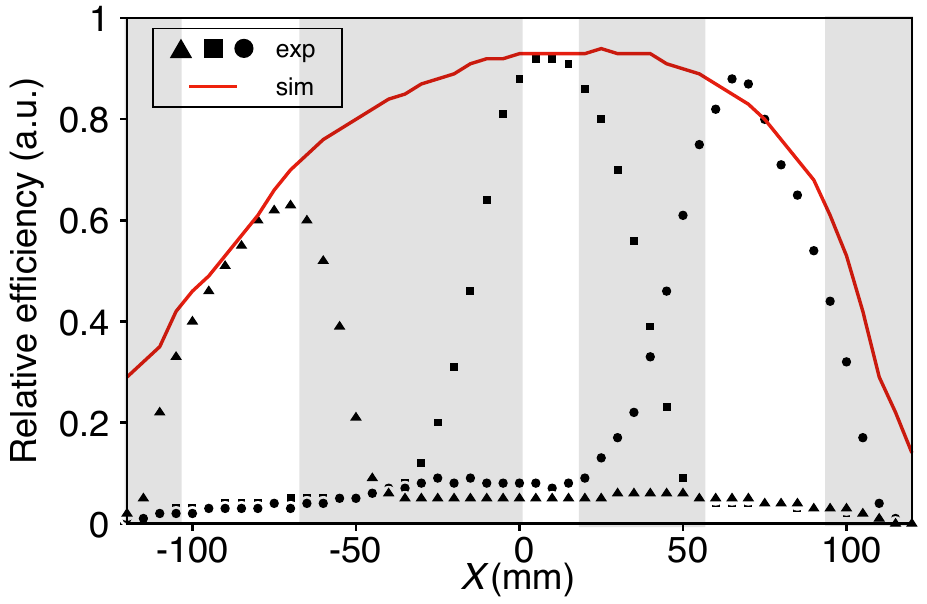}%
\caption{Relative transmission efficiency of the FRS in the high-resolution mode from TA to S4. 
The black symbols and red line denote the experimental data and simulation results for CD$_2$-target, respectively. 
Gray-shaded areas are excluded in the comparison due to nonuniform deuteron distribution.
}
\label{fig:tran}
\end{figure}
Due to the large momentum dispersion mode, the FRS momentum acceptance is non-uniform, requiring transmission efficiency corrections when subtracting $^{11}$C components and determining ($p$,$d$) cross sections. 
We performed Monte Carlo simulations using MOCADI~\cite{Iwasa1997} with FRS optics from GICOSY~\cite{Makino2006} to model deuteron transport from TA to S4, incorporating beam profile, multiple scattering in the target, and beamline geometry. 
The simulations reproduced key optical properties -- the phase space distributions, the dispersion and positions of the focal plane. 
A three-point magnetic rigidity scan was performed in CD$_2$-target measurements at 403 MeV to widen the momentum coverage of elastic $pd$ scattering, as shown by the black symbols in Fig.~\ref{fig:tran}. 
To better reproduce the experimental data, we multiplied the simulated data by a phenomenological quadratic distribution, and obtained the final distribution (red line).
The difference between the corrected and original simulated data is taken as the systematic uncertainty.
Similar simulations were performed at other energies. 

\begin{figure}
\includegraphics[width=0.4\textwidth]{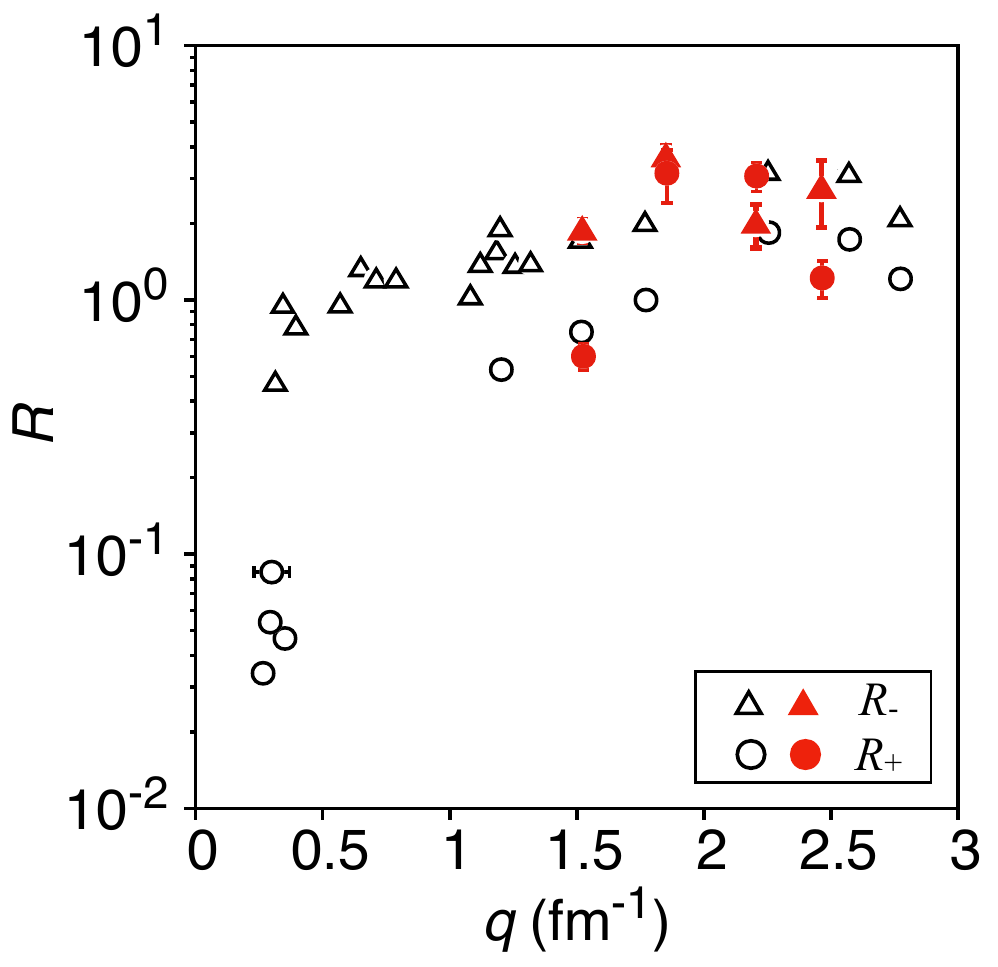}%
\caption{
Cross-section ratios $R_{-}$ (triangles) and $R_{+}$ (circles) versus momentum transfer $q$. 
Filled and open symbols indicate the results of present work at 0$^\circ$ and existing forward-angle data, respectively. See text for details.
}
\label{fig:Rplot}
\end{figure}
To investigate possible tensor correlations in $^{16}$O, we follow the previous work~\cite{Ong2013} and determine the ratios 
%of excited-state transitions relative to the ground-state transition as
$R_{-} = \sigma(^{15}\textrm{O}_{6.2})/\sigma(^{15}\textrm{O}_{\textrm{g.s.}})$ and 
$R_{+} = \sigma(^{15}\textrm{O}_{5.2})/\sigma(^{15}\textrm{O}_{\textrm{g.s.}})$,
where $\sigma (^{15}\text{O}_i)$ denotes the cross section of the transition to a certain final state $i$.
%The ratios were determined 
%after correcting for the FRS transmission %efficiency. 
%The ratios 
$R_{-}$ and $R_{+}$ are denoted in Fig.~\ref{fig:Rplot} by triangles and circles, respectively.
The horizontal axis shows the momentum transfer ($q$), which is the average momentum transfers of the ($p$,$d$) reactions populating the two states. 
%In most cases, the difference in the two momentum transfers is smaller than the symbols, and thus negligible.
Its uncertainty is defined as half the interval between the two corresponding momentum transfers.
The filled symbols are the ratios from this work.
The vertical error bars indicate the overall uncertainty, which includes the statistical and systematic uncertainties.
The systematic uncertainty is attributed mainly to the uncertainty in the analysis procedures involving the spectral fitting and the C target spectrum subtraction (ranging from a few percent to about 20\%), and the estimation of the transmission efficiency ($\leq$ 10\%); the larger of these two contributions is adopted as the systematic uncertainty.
The open symbols represent the experimental data from the literature, where the deuterons are measured at finite scattering angles.
To avoid complication from reaction mechanism, we considered only data at deuteron-scattering angles $\leq$15$^\circ$ (in the center-of-mass system), obtained at various incident proton energies: 30~\cite{Chant1967}, 25--45~\cite{Snelgrove1969}, 65~\cite{Roos1975}, 100~\cite{Lee1967}, 200~\cite{Abegg1989}, 198, 295 and 392~\cite{Ong2013}, and 800 MeV~\cite{Smith1984}.
Uncertainties are included where available.

The present data are in good agreement with the previous results~\cite{Ong2013} at $q \simeq$1.5 fm$^{-1}$. 
The ratio $R_-$ changes gradually, varying by only a factor of 5 -- less than an order of magnitude -- across the range $0.3 \leq q \leq 2.7$ fm$^{-1}$. 
This behavior reflects population of both states through neutron pickup from the $p$-shell, whose wave functions are predominantly zero-particle-zero-hole (0p-0h) configurations and thus lack significant high-momentum components.
In contrast, the ratio $R_{+}$ shows a dramatic increase by two orders of magnitude.
Since the conventional shell model cannot account for positive-parity states, the observed rapid increase aligns with the expectation that tensor correlations generate 2p-2h states, which include high-momentum components.
A distinct peak at around 2 fm$^{-1}$ in $R_{+}$ is observed exclusively in the 0$^\circ$ data, resembling the relative enhancement due to the tensor interactions predicted by various theoretical models~\cite{Schiavilla2007,Neff2003,Alvioli2008}.

In previous work~\cite{Ong2013}, a toy model using Gaussian basis wave functions for $^4$He, constructed within TOSM as detailed in Ref.~\cite{Myo2007}, was applied to calculate the momentum distribution ratio.
% Although the momentum distributions (MD) of nucleons in $^{16}$O have been predicted both with and without tensor interactions using state-of-the-art theoretical frameworks, these predictions have not been made on an orbital-by-orbital basis.
% To deduce the MD of neutrons in different orbitals, we assume that the full momentum distribution ($n_{\textrm{F}}(k)$) consists mainly of central contributions ($n_{\textrm{C}}(k)$) and tensor contributions ($n_{\textrm{T}}(k)$):
This model describes the behavior of $R_{+}$ reasonably well, but fails to account for the peak at around 2 fm$^{-1}$ revealed in the present work.

We compare $R_{+}$ with state-of-the-art theoretical calculations incorporating tensor interactions in $^{16}$O~\cite{Pieper1992,Alvioli2008,Fabrocini2001,Neff2003}.
While these models present the general features of nucleon momentum distributions in $^{16}$O with and without tensor correlations,
they currently lack orbital-by-orbital momentum distributions.
This limitation prevents direct quantitative comparison with present data.
We therefore assume that the momentum distribution is composed of two different components, namely central part ($n_{\textrm{C}}(k)$) and tensor part ($n_{\textrm{T}}(k)$).
The full momentum distribution ($n_{\textrm{F}}(k)$) is written as:
\beq
&& n_{\textrm{F}}(k) = n_{\textrm{C}}(k) + n_{\textrm{T}}(k),
\eeq
where $k$ denotes the internal momentum of neutrons.
% In the main configuration of $^{16}$O, because the contributions of tensor part are expected to be negligibly small, the MD of neutrons in the $p$-shell (corresponding to the negative-parity final state) is taken as:
Then, the momentum distributions of negative parity state ($n_{-}(k)$, the ground state in the present case) and positive parity state ($n_{+}(k)$, the 5/2$^+$ excited state) can be written as:
\beq
% \begin{aligned}
n_{-}(k) &=& a_{-} n_{\textrm{C}}(k) + b_{-} n_{\textrm{T}}(k), \\
n_{+}(k) &=& a_{+} n_{\textrm{C}}(k) + b_{+} n_{\textrm{T}}(k).
% \approx a_{-} n_{\textrm{C}}(k) ,~    
% \end{aligned}
\eeq
% where $a_{-}$ and $b_{-}$ denote the contributions of the central and tensor parts, respectively. 
where $a_{-}$, $a_{+}$, $b_{-}$ and $b_{+}$ are unknown amplitudes associated with the observed states.
% For the MD of neutrons in the $sd$-shell (corresponding to the positive-parity final state), due to the 2p-2h configuration mixing induced by tensor interactions, $n_{+}(k)$ is written as:
% \beq
% &&n_{+}(k) = a_{+} n_{\textrm{C}}(k) + b_{+} n_{\textrm{T}}(k).
% \eeq
% The ratio of MDs of neutrons in the $sd$- and $p$-shell is then written as:
The ratio $R_{+}$ is expressed as: $R_{+}(k) = n_+(k)/n_-(k)$.
%\beq
%R_{+}(k) &=& \frac{n_+(k)}{n_-(k)} .~   \label{eq_R}
% \approx u+v \frac{n_{\textrm{T}}(k)}{n_{\textrm{C}}(k)},
%\eeq
Since the negative parity states are mainly formed by 0p-0h configurations, where the contribution of the tensor interactions is considered negligible ($b_{-} \sim 0 $),
$R_{+}(k)$ can be reduced to:
\beq
R_{+}(k) &\approx& u+v \frac{n_{\textrm{T}}(k)}{n_{\textrm{C}}(k)},
\eeq
where $u=a_+/a_-$ and $v=b_+/a_-$.

To derive $n_{\textrm{T}}(k)/n_{\textrm{C}}(k)$, we use the MDs predicted by various models with distinct interactions.
We adopt the MD without tensor interactions as $n_{\textrm{C}}(k)$ and the one with tensor interactions as $n_{\textrm{F}}(k)$.
Taking the variational Monte Carlo (VMC) calculations with cluster expansion (Fig.~3, Ref.~\cite{Pieper1992}) as an example, we extracted the MD from the dot-dashed line corresponding to the ``Jastrow" wave function (with only central correlations) as $n_{\textrm{C}}(k)$, and from the solid line representing the ``Jastrow + two-body" as $n_{\textrm{F}}(k)$, since the two-body cluster expansion primarily accounts for tensor correlations.
The model ratio then becomes ${n_{\textrm{T}}}/{n_{\textrm{C}}} = {n_{\textrm{F}}}/{n_{\textrm{C}}} - 1$.
We assume $n_{\textrm{T}}/n_{\textrm{C}} \approx 0$ at low momentum, where the tensor contributions are negligible compared with the central part.
The renormalization of the central momentum distribution ${n_{\textrm{C}}}$ is omitted in this analysis.
The MDs calculated with and without tensor interactions are both normalized to the nucleon number ($A=16$) in $^{16}$O. 
While the central momentum distribution should ideally be renormalized to account for tensor-correlated nucleons,
% (with a minimum scaling factor of 14/16 for two-particle correlations), 
this effect is negligible compared to the observed orders-of-magnitude variation in $R_+$.
Therefore, neglecting renormalization does not affect the physical interpretation of our results.
    
In ($p$,$d$) reactions within the pickup-dominant region, 
% $q \approx k$, 
the momentum transfer ($q$) is equal to the internal momentum of neutron ($k$).
Hence, experimental $R_{+}(q)$ can be compared with $R_{+}(k)$, provided $u$ and $v$.
Due to the absence of orbital-by-orbital calculations for $^{16}$O including high-momentum components, we fix $u$ and $v$ using the experimental $R_{+}(q)$.
At $q\sim0$ fm$^{-1}$, $R_{+}\approx u$, 
and thus $u=R_{+}^{\textrm{min}}=0.055$, considering the mean of the 22--45-MeV data~\cite{Snelgrove1969}.
The maximum value $R_{+}^{\textrm{max}}\approx 3.1$ from the present work is used to determine $v=0.29$ for VMC.
 
\begin{figure}
\includegraphics[width=0.40\textwidth]{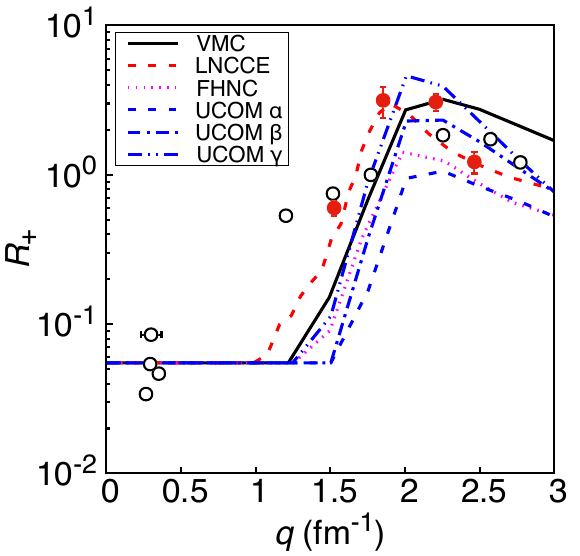}%
\caption{Comparison of the cross-section ratios between experimental data and theoretical calculations. Filled symbols indicate the experimental data with the same definition in Fig.~\ref{fig:Rplot}. Lines represent the ratios deduced from theoretical calculations~\cite{Alvioli2008,Pieper1992,Fabrocini2001,Neff2003}. See text for further details.}
\label{fig:discu}
\end{figure}

% Since the tensor contributions are expected to be small in $R_-$, we consider only $R_+$ in the following discussion. 
Figure~\ref{fig:discu} shows the comparison of the experimental and theoretical $R_+$ ratios.
The definition of symbols follows that of Fig.~\ref{fig:Rplot}. 
Results of the VMC~\cite{Pieper1992}, the linked and number conserving cluster expansion (LNCCE)~\cite{Alvioli2008}, and Fermi hypernetted chain (FHNC)~\cite{Fabrocini2001} models are shown by the black-solid, red-dashed and magenta-dotted lines, respectively. 
For the LNCCE and FHNC models, we used the ``Full" and ``Central" lines (Fig.~2, Ref.\cite{Alvioli2008}), and the ``$f_6$" and ``Jastrow" lines (Fig.~2, Ref.~\cite{Fabrocini2001}) as $n_{\textrm{F}}(k)$ and $n_{\textrm{C}}(k)$, respectively. 
The blue lines show the ratios deduced using the distributions in Fig. 29 of Ref.~\cite{Neff2003} obtained with the unitary correlation operation method (UCOM); the line of ``radial" is used as $n_{\textrm{C}}(k)$ and the lines of three different range parameters for the tensor interactions -- $\alpha$, $\beta$ and $\gamma$ -- are used as $n_{\textrm{F}}(k)$, and the results are shown by the dash, dot-dashed, double-dot-dashed lines, respectively. 
The same $u$ and $v$ values obtained by VMC were adopted for all other models.

The experimental data and theoretical models exhibit good agreement, showing a peak in the ratio at momentum transfer near 2 fm$^{-1}$, highlighting the dominant role of tensor contributions in the momentum-transfer dependence of the ratio. 
As momentum increases, the experimental ratio declines,
% and approaches unity at $q \geq$ 2.5 fm$^{-1}$
consistent with the findings from the recent electron-scattering study~\cite{Schmidt2020}. 
However, while theory predicts a sharp decline after the peak as momentum decreases, experimental ratios remain high down to lower momenta, dropping only at very small amount. 
This unexpected behavior was also reported in Ref.~\cite{Miki2013}, which studied ($S$=1,~$T$=0) to ($S$=0,~$T$=1) proton-neutron pairs in $^4$He. 
Deviation is also observed in the experimental data from Ref.~\cite{Ong2013}, notably around $q\sim 1.2$ fm$^{-1}$. 
Since these experiments, including the data from Ref.~\cite{Snelgrove1969}, were conducted near 10$^\circ$, the discrepancy may stem from distortion effects.
Indeed, DWBA calculations at 45 MeV~\cite{Snelgrove1969,Pang_private} suggest the ratio rises sharply at small angles, indicating a much larger $u$.
It is therefore desirable to obtain not only 0$^\circ$ experimental data at low energy, but also extend theoretical studies to include rigorous orbital-by-orbital calculations incorporating tensor interactions and high-momentum components to provide more exclusive $v$ value.

%\section{Summary} 
The $^{16}$O($p$,$d$)$^{15}$O reaction was studied at proton energies of 403, 604, 907, and 1209 MeV with deuterons detected at 0$^\circ$. 
The cross-section ratios for populating $^{15}$O excited states versus the ground state were systematically studied as a function of momentum transfer. 
A pronounced enhancement near 2 fm$^{-1}$ was observed for positive-parity state, attributed to tensor interactions that introduce high-momentum neutrons via two-particle-two-hole configuration mixing. 
Theoretical models incorporating tensor interactions show good agreement with experimental data, reproducing the peak structure at around 2 fm$^{-1}$ for the $5/2^+$-to-ground-state ratio. 
These results demonstrate that high-momentum neutron-pickup reactions at 0$^\circ$ are a sensitive probe for high-momentum components driven by the short-range nature of tensor interactions. 
The present findings open new avenues for investigating effects of tensor interactions in nuclei. Performing ($p$,$d$) reactions in inverse kinematics using intermediate- to high-energy radioactive beams at FAIR~\cite{FAIR2025} and HIAF~\cite{Zhou2022} to study exotic nuclei along isotopic and isotonic chains would be particularly valuable for elucidating the role of tensor interactions in shell evolution~\cite{Otsuka2005}, intruder orbitals~\cite{Myo2008} and dense nuclear matter~\cite{Dickhoff2004}.

\begin{acknowledgments}
The authors acknowledge the support from various GSI departments, particularly the accelerator group and operators, the GSI target laboratory, and the engineers of the FRS group. H. J. O. thanks M. Lyu and K. Ogata for inspiring discussion, and acknowledges the support of CAS President's International Fellowship Initiative.
X. W. acknowledges the support of scholarships from the Graduate School of Science, Osaka University and the Daiyukyo Foundation of International Exchange Assistance and Research.
This work was supported by the Hessian Ministry for Science and Art (HMWK) through the LOEWE Center HICforFAIR, and by Justus-Liebig-Universität Gießen and GSI under the JLU-GSI strategic Helmholtz partnership agreement.
Partial support was also provided by the JSPS Grant-in-Aid for Scientific Research No. 23224008, National Natural Science Foundation of China (12175009, 12175280), the International Partnership Program of the Chinese Academy of Sciences (016GJHZ2023063GC), and Major Science and Technology Projects in Gansu Province (24GD13GA005).

\end{acknowledgments}
\bibliography{s436letters}

\end{document}